# Cluster mass estimates from weak lensing

Matthias Bartelmann[1,2]

[1]Harvard-Smithsonian Center for Astrophysics, 60 Garden Street, Cambridge, MA 02138, USA;
[2]Max-Planck-Institut für Astrophysik, Postfach 1523, D–85740 Garching, FRG

*Abstract.* The reliability of cluster lens reconstruction techniques based on weak lensing is studied in terms of the accuracy of their reproduction of the total cluster mass as a function of distance from the cluster center. To do so, a variety of reconstruction algorithms is applied to synthetic lensing data created using a sample of 60 numerically modeled clusters, and the mass reconstruction is compared to the known deflector mass. The results can be summarized as follows: (1) Reconstruction algorithms which require integrations extending over the entire real plane yield unreliable results, because they give rise to boundary effects which are hard to control; mass overestimates are more likely and more substantial in this case than underestimates. (2) Reconstruction techniques which avoid these boundary effects yield reliable lower bounds to the cluster mass. The tightness of such bounds depends on the size of the field, which can be extended synthetically to improve the results considerably. For the sample of numerical cluster models, the best lower bounds, achieved by combining synthetic field extension with non-linear, finite-field reconstruction, decline from 100% to 80% of the true cluster mass going from the cluster center to an angular distance of $2\rlap{.}'5$. The 80% error bars of the lower mass bounds are $\pm 10\%$ to $15\%$.

## 1 Introduction

It was realized by R. Webster as early as in 1985 that galaxy clusters could coherently distort images of background galaxies, and that information about the mass distribution in these clusters could be inferred from that distortion. After the detection of the faint blue galaxy population (Tyson 1988), first observational evidence for coherent distortions around galaxy clusters was found (Tyson, Valdes & Wenk 1990), and then attempts were made to systematically fit parametrized cluster mass distributions to the observed image distortions (Kochanek 1990, Miralda-Escudé 1991). Kaiser & Squires (1993) were the first to notice that, in the weak lensing regime, parameter-free cluster-lens inversions are possible on the basis of the observed image ellipticities. While Kaiser & Squires first suggested the application of their method to statistical cluster samples rather than to individual clusters except for very rich ones, several applications of their inversion technique to real data proved the feasibility of their method (Fahlman et al. 1994, Smail et al. 1994, Kaiser et al. 1994b,c).

While the overall agreement between the reconstructed mass maps and independent information derived from the galaxy- or the X-ray distribution demonstrated that the method worked successfully in principle, some intriguing puzzles were encountered. First, Fahlman et al. (1994) reported a total mass derived from lensing by the cluster MS 1224+20 three times as large as the cluster's virial mass within an aperture of radius $2.'76$ around the cluster center. This finding was qualitatively confirmed by Carlberg, Yee & Ellingson (1994), who found the lensing mass $2.5 \pm 1.1$ times larger than the virial mass. Fitting mass profiles to the weak shear observed in the field of $Cl\,0024 + 1654$, Bonnet, Mellier & Fort (1994) also found a surprisingly large mass for this cluster. Conversely, Squires (1994) reported that the cluster-inversion algorithm failed to detect any significant mass at all in the field of A 2163, which is particularly disturbing because this is the hottest and most luminous X-ray cluster in the sky.

Recent theoretical studies aimed at removing sources of systematic error from the most straightforward cluster-reconstruction algorithm. As mentioned by Kaiser & Squires (1993) and discussed by Schneider & Seitz (1994), the surface-density map derived from coherent image distortions is invariant to a global linear transformation which expresses the fact that a matter sheet of constant surface density does not change the distortion pattern at all. This degeneracy can only be broken if independent information on the absolute surface mass density can be obtained, e.g., by observing the magnification of galaxy images behind clusters (cf. Broadhurst, Taylor & Peacock 1994, Narayan & Bartelmann 1994). The original method by Kaiser & Squires does only apply in the weak-lensing limit, i.e., if convergence and shear are small compared to unity. Two problems arise if this limit is not met, first, the observed image distortions then measure a combination of shear and convergence rather than the shear alone as in the case of weak lensing, second, the possible formation of critical lines by strong lenses causes a further degeneracy because then a given image distortion can arise outside and inside the critical curve, and they cannot be distinguished because the parity of images cannot be observed. Both these problems were solved by Seitz & Schneider (1994), who suggested an iterative algorithm capable of simultaneously reconstructing weak- and strong-lensing regimes (see also Kaiser 1994). Another source of systematic error is caused by the fact that the most simple variant of the cluster-reconstruction algorithm requires a convolution to be performed on the entire real plane, while actual observational data are necessarily restricted to finite fields. If applied to finite fields, however, the cluster inversion yields a mass distribution with vanishing total mass in the field. A method to remove this systematic boundary effect was first suggested by Schneider (1994), and Kaiser et al. (1994c) discussed variants of this scheme.

It is evident from the quoted references that this field of astrophysics is evolving at a remarkable rate. This reflects the considerable interest attracted by the cluster-inversion technique, which is due to the facts that gravitational lensing provides essentially the only method to learn about the true mass distribution of clusters without referring to any equilibrium or stability assumptions, and that it was proved feasible to observe the weak lens effect by clusters with sufficient signal-to-noise ratio (see Kaiser, Squires & Broadhurst 1994b for a description and discussion of methods of data analysis).

This paper pursues the following question: Given "ideal" observational data, that is to say, observational data which have been cleaned of systematic observational errors due to the telescope or the CCD, and which are not contaminated by galaxies belonging to the cluster lens itself, what is the accuracy with which we can hope to reconstruct from weak lensing the total cluster mass as a function of distance from the cluster center? To answer this question, the weak lens effect of a sample of 60 numerically modeled galaxy clusters is simulated, and from the simulated background galaxy ellipticities the clusters are reconstructed using a variety of reconstruction techniques. To set the stage, I review in Sect. 2 the reconstruction algorithms whose application to the numerically modeled galaxy clusters is described in Sect. 3. In Sect. 4, I briefly discuss the influence of seeing and the unknown source redshift distribution, and the results are summarized and discussed in Sect. 5.

## 2 Cluster mass estimators

### 2.1 Basic relations

This section reviews a number of different methods to estimate cluster masses from their weak lens effect. All these methods are physically based on the fact that the local properties of a lens, its convergence $\kappa$ and shear $\gamma$, are related through the same scalar potential $\psi$,

$$\psi(\boldsymbol{x}) = \frac{1}{\pi} \int_{\mathbb{R}^2} d^2 x' \ \kappa(\boldsymbol{x}') \ln |\boldsymbol{x} - \boldsymbol{x}'| \ , \tag{2.1}$$

where $\boldsymbol{x}$ is a two-dimensional position vector on the observer's sky. In the weak-lensing limit, the distortion of background galaxy images is determined by the shear, while the convergence is just the surface mass density in units of its critical value. Starting from the potential $\psi(\boldsymbol{x})$, the convergence is determined by the two-dimensional version of Poisson's equation, or

$$\kappa(\boldsymbol{x}) = \frac{1}{2} \Delta \psi = \frac{1}{2} [\psi_{,11}(\boldsymbol{x}) + \psi_{,22}(\boldsymbol{x})] \ , \tag{2.2}$$

where an index $j$ preceded by a comma denotes partial differentiation with respect to $x_j$. The components of the shear are likewise given by

$$\gamma_1(\boldsymbol{x}) = \frac{1}{2} [\psi_{,11}(\boldsymbol{x}) - \psi_{,22}(\boldsymbol{x})] \ , \quad \gamma_2(\boldsymbol{x}) = \psi_{,12}(\boldsymbol{x}) \ . \tag{2.3}$$

Equations (2.1) through (2.3) provide the complete basis for all cluster reconstruction techniques. For a detailed discussion of their derivation and their properties, see Schneider, Ehlers & Falco (1992, Chap. 5).

It will later prove useful to combine the shear components into the complex shear $\gamma$,

$$\gamma(\boldsymbol{x}) \equiv \gamma_1(\boldsymbol{x}) + \mathrm{i}\gamma_2(\boldsymbol{x}) \ . \tag{2.4}$$

Kaiser & Squires (1993) used the Fourier transform to show that $\kappa$ and $\gamma$ are related by the convolution

$$\kappa(\boldsymbol{x}) = -\frac{1}{\pi} \int_{\mathrm{I\!R}^2} d^2 x' \ \mathrm{Re}\left[\mathcal{D}(\boldsymbol{x}-\boldsymbol{x}')\gamma^*(\boldsymbol{x}')\right] \ , \tag{2.5}$$

where the superscribed asterisk denotes complex conjugation, and the complex convolution kernel is given by

$$\mathcal{D}(\boldsymbol{x}) = \frac{x_1^2 - x_2^2 + 2\mathrm{i}x_1 x_2}{x^4} \ . \tag{2.6}$$

Equation (2.5) can be verified by inserting Eq. (2.1) into Eq. (2.3), obtaining a relation between $\gamma$ and $\kappa$, and then inserting into that relation Eq. (2.5).

*2.2 Linear cluster reconstruction*

In the limit of weak lensing, $|\gamma| \ll 1$ and $\kappa \ll 1$, the local ellipticity of galaxy images is a measure for the local shear $\gamma(\boldsymbol{x})$. The usual way to proceed is to measure the surface-brightness quadrupole $Q_{jk}(\boldsymbol{x})$ of each galaxy image,

$$Q_{jk} \equiv \frac{\int d^2 x \ x_j x_k I(\boldsymbol{x})}{\int d^2 x \ I(\boldsymbol{x})} \ , \tag{2.7}$$

where $I(\boldsymbol{x})$ is the surface-brightness distribution of the image, and to form the complex (or two-component) ellipticity $\chi$,

$$\chi = \chi_1 + \mathrm{i}\chi_2 \equiv \frac{Q_{11} - Q_{22} + 2\mathrm{i}Q_{12}}{Q_{11} + Q_{22}} \ . \tag{2.8}$$

If the lens is weak, and if the lensed sources were intrinsically circular, the image ellipticity would be twice the shear, $\chi = 2\gamma$. In the most straightforward linear reconstruction algorithm, the ellipticity $\chi$ is therefore taken as a local estimator of $2\gamma$ and inserted into Eq. (2.5) to yield

$$\kappa(\boldsymbol{x}) \simeq -\frac{1}{2\pi n} \sum_{j=1}^{N} \mathrm{Re}\left[\mathcal{D}(\boldsymbol{x}-\boldsymbol{x}_j)\chi_j^*(\boldsymbol{x}_j)\right] \ , \tag{2.9}$$

where the area integral has been replaced by a sum extending over $N$ galaxies, whose number density is $n$.

In practice, a straightforward application of Eq. (2.9) is impaired by the fact that the lensed galaxies are intrinsically elliptical. The randomness of these ellipticities implies that the noise introduced by them has a white spectrum (see Kaiser & Squires 1993). The high-frequency contribution to the noise is therefore divergent, and this divergence must be suppressed by introducing a low-pass filter into Eq. (2.9). Intuitively, the divergent noise is caused by the singularity in the convolution kernel $\mathcal{D}(\boldsymbol{x})$ for $|\boldsymbol{x}| \to 0$, which adds divergent weight to the galaxy images very close to the point under consideration. Introducing the low-pass filter $W(\boldsymbol{x}; \Delta x)$ modifies Eq. (2.9) to read

*Cluster mass estimators* 5$$\kappa(\boldsymbol{x}) \simeq -\frac{1}{2\pi n} \sum_{j=1}^{N} W(\boldsymbol{x} - \boldsymbol{x}_j; \Delta x) \, \text{Re}\left[\mathcal{D}(\boldsymbol{x} - \boldsymbol{x}_j)\chi_j^*(\boldsymbol{x}_j)\right] \; . \tag{2.10}$$

Since the only purpose of the filter is to suppress the signal for $\boldsymbol{x} \to \boldsymbol{x}'$, any isotropic function $W(\boldsymbol{x}; \Delta x)$ is sufficient which tends to zero for $|\boldsymbol{x}| \to 0$ on a reasonable scale $\Delta x$, and approaches unity for $|\boldsymbol{x}| \gg \Delta x$. An appropriate choice is (Seitz & Schneider 1994)

$$W(\boldsymbol{x}, \Delta x) = 1 - \left(1 + \frac{x^2}{2\Delta x^2}\right) \exp\left(-\frac{x^2}{2\Delta x^2}\right) \; , \tag{2.11}$$

where $\Delta x$ has to be chosen according to the spatial galaxy density, see below. Equation (2.10), supplemented by some choice of the smoothing kernel $W$, provides a first simple estimator for the convergence $\kappa(\boldsymbol{x})$ at a point $\boldsymbol{x}$.

*2.3 Smoothing prior to reconstruction*

A reconstruction following Eq. (2.10) avoids the noise divergence caused by the intrinsic galaxy ellipticities, but still suffers from the randomness of the galaxy positions, which introduces "shot noise" into the reconstruction. A viable way to cure that is to smooth the observed data before applying the reconstruction algorithm to them, i.e., to construct a smooth ellipticity map from the data before inserting them into Eq. (2.5). For instance, such an ellipticity map can be determined at the points of a regular grid $\boldsymbol{x}_{jk}$ by the weighted average

$$\bar{\chi}(\boldsymbol{x}_{jk}) = \frac{\sum_{l=1}^{N} w_l \chi(\boldsymbol{x}_l)}{\sum_{l=1}^{N} w_l} \; , \tag{2.12}$$

where Gaussian weights $w_l$ appear appropriate,

$$w_l = \exp\left[-\frac{(\boldsymbol{x}_{jk} - \boldsymbol{x}_l)^2}{\Delta x^2}\right] \; . \tag{2.13}$$

Since Eq. (2.12) is a weighted average over the ellipticity data, it also reduces the noise introduced by the intrinsic galaxy ellipticities. Inserting $\bar{\chi}(\boldsymbol{x})$ into Eq. (2.5), we arrive at the alternative linear reconstruction equation

$$\kappa(\boldsymbol{x}) \simeq -\frac{a^2}{2\pi} \sum_{j,k} \text{Re}\left[\mathcal{D}(\boldsymbol{x} - \boldsymbol{x}_{ij})\bar{\chi}^*(\boldsymbol{x}_{jk})\right] \; , \tag{2.14}$$

where $a$ is the separation of grid points.

An obvious choice for the smoothing length $\Delta x$, which occurs in the smoothing kernel $W$ of Eq. (2.11) and in the Gaussian weights $w_l$ of Eq. (2.13), is the average distance between galaxies,

$$\Delta x_0 = \frac{1}{\sqrt{n\pi}} \; ; \tag{2.15}$$

then, there will on average be one galaxy image within a smoothing disk. This is sufficient if the measured "signal" is strong, that is to say, if locally $|\chi| \to 1$. It should suffice in that case to use the ellipticity of one or very few galaxies to estimate the local shear. If the signal is "weak", the smoothing scale should be extended such that a smoothing disk contains a larger number of galaxies. However, what should be considered a weak signal depends on the intrinsic ellipticity distribution of the sources. To be specific, let us assume an intrinsic ellipticity distribution of the form

$$p_{\rm s}(|\chi_{\rm s}|) = \frac{\exp\left(-|\chi_{\rm s}|^2/\chi_{\rm r}^2\right)}{\pi\chi_{\rm r}^2\left[1-\exp(-1/\chi_{\rm r}^2)\right]}\;,\quad |\chi_{\rm s}|\in[0,1]\;, \tag{2.16}$$

where $\chi_{\rm r}$ measures the scatter in the intrinsic ellipticity distribution. Clearly, the signal $|\chi|$ should be considered weak when $|\chi|\to\chi_{\rm r}$. Therefore, I adapt $\Delta x$ to the strength of the signal such that

$$\frac{\Delta x}{\Delta x_0} \simeq \begin{cases} 1 & \text{for} \quad |\chi|\to 1 \\ 3 & \text{for} \quad |\chi|\to\chi_{\rm r} \end{cases}; \tag{2.17}$$

this means that a smoothing disk contains on average about ten galaxies when the signal is weak, and one galaxy when the signal is strong. A simple linear interpolation scheme which fulfils these conditions is

$$\Delta x = \Delta x_0 \left(3 - 2\frac{|\chi|-\chi_{\rm r}}{1-\chi_{\rm r}}\right)\;. \tag{2.18}$$

To clarify the notation, 'linear reconstruction' without or with smoothing in the following refers to the reconstruction equations (2.10) or (2.14), respectively, regardless of the fact that Eq. (2.10) already involves low-pass filtering the data.

### 2.4 Iterative nonlinear reconstruction

The linear reconstruction formulae (2.10) and (2.14) are valid in the limit of weak lensing. If this limit does not apply, several complications arise. First of all, Schneider & Seitz (1994) have emphasized that the only observable quantity accessible through the ellipticities of the galaxy images is the distortion $\delta$,

$$\delta = \frac{2\gamma(1-\kappa)}{(1-\kappa)^2+|\gamma|^2}\;, \tag{2.19}$$

which is related to the shear by

$$\gamma = \frac{1-\kappa}{\delta^*}\left[1\pm\sqrt{1-|\delta|^2}\right]\;; \tag{2.20}$$

see also Miralda-Escudé (1991). [The fact that only the combination $\delta$ of $\kappa$ and $\gamma$ is observable via ellipticities is most easily verified mapping a circular source with the inverse Jacobian of the lens mapping, $\mathcal{A}^{-1} = \mathrm{diag}[(1-\kappa-\gamma)^{-1},(1-\kappa+\gamma)^{-1}]$, and determining the ellipticity of the image.] The choice of the sign in Eq. (2.20) introduces a local degeneracy which is physically caused by the fact that a given distortion of the galaxy images can be produced on either side of the critical curve of a lens, because the parity of the images cannot be observed. The correct choice for the sign is the negative sign of the determinant of the local Jacobian matrix $\mathcal{A}$. Two problems therefore arise immediately in the case of strong lensing, first, the distortion $\delta$ has to be inferred from the ellipticity measurements, second, the sign in Eq. (2.20) has to be chosen correctly. Both these problems have recently been solved by Seitz & Schneider (1994).

They showed that the condition that the intrinsic galaxy ellipticities should be randomly oriented (i.e., $\chi$ has random phase intrinsically) can be translated into the requirement

$$\sum_{j=1}^{N'} \exp\left[-\frac{(\boldsymbol{x}-\boldsymbol{x}_j)^2}{\Delta x^2}\right] \frac{\delta(\boldsymbol{x})-\chi_j}{1-\mathrm{Re}\left[\delta(\boldsymbol{x})\chi_j^*\right]} = 0 \;, \quad (2.21)$$

where the Gaussian factor was again introduced for smoothing the shot noise. This implicit equation for the distortion $\delta$ can be solved iteratively, starting the iteration with $\delta^{(0)} = \chi$. The sum in Eq. (2.21) extends over all $N'$ galaxies in a patch of the observed field large enough to contain a sufficient number of galaxies and small enough so that the distortion does not change significantly within the patch. Again, the size of the patch can be adapted to the strength of the signal, as described before.

Having solved for the distortion field $\delta$, Eq. (2.20) can be used to obtain an estimate for the shear. Since the convergence and the position of the critical curve are yet unknown, $\kappa = 0$ and $\det \mathcal{A} > 0$ are assumed in a first step, yielding

$$\gamma^{(0)}(\boldsymbol{x}) = \frac{1-\sqrt{1-|\delta(\boldsymbol{x})|^2}}{\delta^*(\boldsymbol{x})} \;. \quad (2.22)$$

The procedure now continues iteratively. In each step of the iteration, Eq. (2.5) is used to determine

$$\kappa^{(n)}(\boldsymbol{x}) = -\frac{a^2}{\pi} \sum_{j,k} \mathrm{Re}\left[\mathcal{D}(\boldsymbol{x}-\boldsymbol{x}_{jk})\gamma^{*(n-1)}(\boldsymbol{x}_{jk})\right] \;, \quad (2.23)$$

and from $\kappa^{(n)}$ and the shear $\gamma^{(n)}$,

$$\gamma^{(n)}(\boldsymbol{x}) = -\frac{a^2}{\pi} \sum_{jk} \mathcal{D}(\boldsymbol{x}-\boldsymbol{x}_{jk})\kappa^{(n)}(\boldsymbol{x}_{jk}) \;, \quad (2.24)$$

the $n$-th iteration of the sign of the Jacobian can be found,

$$\mathrm{sign}\left(\det \mathcal{A}^{(n)}\right) = \mathrm{sign}\left[(1-\kappa^{(n)})^2 - |\gamma^{(n)}|^2\right] \;. \quad (2.25)$$

The next iteration step starts again with the determination of $\gamma$ from Eq. (2.20), inserting the sign and $\kappa$ determined in the previous step. In practice, as long as the critical curve of the lens being reconstructed is not yet known, a further smoothing kernel should be introduced into Eq. (2.23) to smooth over the singularity which occurs in Eq. (2.20) if $\delta \to 0$ when the sign is negative. For an appropriate choice of this kernel, and for further details on this method, I refer to Seitz & Schneider (1994).

The iteration is continued until a reasonable reproduction of the observed distortion field is achieved, which can be estimated by means of a $\chi^2$ comparison between the observed and the reconstructed distortion field, given the intrinsic ellipticity distribution of the galaxy ellipticities. Again, Seitz & Schneider (1994) give a detailed description of the convergence properties of their iterative algorithm.

*2.5 Boundary effects*

The three methods of cluster reconstruction described up to now all involve convolution integrals extending over the entire real plane. Observational data, however, are limited to finite frames. If applied to finite data, the previously described techniques tend to construct spurious negative troughs in the surface-density maps. Several methods have been suggested so far to reduce or remove these effects. The one which is easiest to implement was first mentioned by Seitz & Schneider (1994). They proposed to fit a spherically symmetric, analytic lens model to the distortion map obtained from the data, that is, the parameters of a lens model like

$$\kappa_{\mathrm{model}}(x) = \frac{\kappa_0}{(x^2 + x_{\mathrm{core}}^2)^{p/2}} \quad (2.26)$$

are chosen such as to minimize the mean-square deviation between the distortion fields produced by the model and reconstructed from the data. The justification of this procedure is based on the expectation that the outer parts of clusters tend to become increasingly symmetric with increasing distance from the center. Therefore, the adaptation of the lens model is performed using only the fraction of the field outside a specified radius $x_{\mathrm{fit}}$, which can be on the order of arc minutes. This procedure can be used to artificially extend the field into regions where $\kappa$ and $\gamma$ are well below any measurable value. I refer to this technique in the following as "field extension".

Other more involved methods to remove the boundary effects introduced by the finite size of the field start with Kaiser's (1994) realization that the gradient of $\kappa$ can be written as a linear combination of derivatives of $\gamma$,

$$\nabla \kappa = \begin{pmatrix} \gamma_{1,1} + \gamma_{2,2} \\ \gamma_{2,1} - \gamma_{1,2} \end{pmatrix} , \quad (2.27)$$

so that the difference in $\kappa$ between any two points, $\boldsymbol{x}_0$ and $\boldsymbol{x}$, can be found by integrating the components of $\nabla \kappa$ along arbitrary lines connecting the two points,

$$\kappa(\boldsymbol{x}) = \kappa(\boldsymbol{x}_0) + \int_{\boldsymbol{x}_0}^{\boldsymbol{x}} d\boldsymbol{l} \cdot \nabla \kappa . \quad (2.28)$$

To obtain an estimate of $\kappa(\boldsymbol{x})$ at a given point $\boldsymbol{x}$, one can choose $\boldsymbol{x}_0$ on the boundary of the field, integrate towards $\boldsymbol{x}$, and then average the result over all boundary points. If the boundary is sufficiently distant from the center of the cluster, $\kappa(\boldsymbol{x}_0)$ can safely be assumed to vanish; otherwise, the procedure estimates the difference

$$\kappa(\boldsymbol{x}) - \oint d\phi \, \kappa[\boldsymbol{x}_0(\phi)] , \quad (2.29)$$

where $\boldsymbol{x}_0(\phi)$ parametrizes the boundary of the field. A variant of this technique to reconstruct $\kappa$ was suggested by Schneider (1994). Another way to use Eq. (2.27), which is numerically easier to implement, was suggested by Kaiser et al. (1994c); they proposed to average the starting points $\boldsymbol{x}_0$ of the line integrals over the entire observed field. As shown below, this essentially leads to modification of the convolution kernel of the previous reconstruction formulae.

Suppose the boundary of the field be given by a closed curve $\boldsymbol{c}(\phi)$, parametrized by the continuous parameter $\phi \in [0, 2\pi]$; further, let $R(\phi) \equiv |\boldsymbol{c}(\phi)|$ be the distance of the boundary from the center of the field, and $A$ be the area enclosed by $\boldsymbol{c}$. Without loss of generality, let the target point, i.e. the point where $\kappa$ should be reconstructed, be at the coordinate origin. Then, transforming coordinates $(x_1, x_2)$ to $(t, \phi)$, where

$$x_1 = tR\cos\phi, \quad x_2 = tR\sin\phi, \tag{2.30}$$

radial lines connecting the boundary with the origin are given by $l(t,\phi) = t(\cos\phi, \sin\phi)$ with $t \in [0,1]$, and the average of Eq. (2.28) over $A$ reads

$$\kappa(0) - \bar{\kappa} = \\ -\frac{1}{A} \int_0^{2\pi} R^3(\phi) d\phi \int_0^1 t\, dt \int_0^t dt' \left[ \cos\phi(\gamma_{1,1} + \gamma_{2,2}) + \sin\phi(\gamma_{2,1} - \gamma_{1,2}) \right]. \tag{2.31}$$

$\bar{\kappa}$ denotes the average value of $\kappa$ within $A$. Since the derivatives of $\gamma_j$ have to be taken along the paths, they depend on $t'$, and therefore the $t$-integration can be performed. Renaming $t'$ to $t$ afterwards, we obtain

$$\kappa(0) - \bar{\kappa} = \\ -\frac{1}{2A} \int_0^{2\pi} R^3(\phi) d\phi \int_0^1 dt (1 - t^2) \left[ \cos\phi(\gamma_{1,1} + \gamma_{2,2}) + \sin\phi(\gamma_{2,1} - \gamma_{1,2}) \right]. \tag{2.32}$$

Transforming back to cartesian coordinates (2.30), and inserting $\cos\phi = x_1/x$, $\sin\phi = x_2/x$, Eq. (2.32) can be written

$$\kappa(0) - \bar{\kappa} = \frac{1}{2A} \int dx_1 \int dx_2 \left(1 - \frac{R^2}{x^2}\right) (x_1 \gamma_{1,1} + x_2 \gamma_{2,1} + x_1 \gamma_{2,2} - x_2 \gamma_{1,2}). \tag{2.33}$$

Using the relations

$$\frac{\partial}{\partial x_1}\left(1 - \frac{R^2}{x^2}\right) = \frac{2R}{x^4}\left(Rx_1 + \dot{R}x_2\right) \\ \frac{\partial}{\partial x_2}\left(1 - \frac{R^2}{x^2}\right) = \frac{2R}{x^4}\left(Rx_2 - \dot{R}x_1\right), \tag{2.34}$$

where $\dot{R}$ denotes $(dR/d\phi)$, Eq. (2.33) can be integrated by parts to yield

$$\kappa(0) - \bar{\kappa} = \\ -\frac{1}{A} \int d^2x \left[ \frac{R^2}{x^2}(\gamma_1 \cos 2\phi + \gamma_2 \sin 2\phi) + \frac{R\dot{R}}{x^2}(\gamma_1 \sin 2\phi - \gamma_2 \cos 2\phi) \right]. \tag{2.35}$$

Shifting the target point into an arbitrary position $\boldsymbol{x}$ within the boundary $\boldsymbol{c}(\phi)$, we finally obtain

$$\kappa(\boldsymbol{x}) - \bar{\kappa} = \\ -\frac{1}{A} \int_A d^2x' \left\{ R^2 \operatorname{Re}[\mathcal{D}(\boldsymbol{x} - \boldsymbol{x}')\gamma^*(\boldsymbol{x}')] - R\dot{R} \operatorname{Im}[\mathcal{D}^*(\boldsymbol{x} - \boldsymbol{x}')\gamma(\boldsymbol{x}')] \right\}. \tag{2.36}$$

Apart from changes of notation, Eq. (2.36) agrees with Eq. (7) of Kaiser et al. (1994c). Note that (2.36) reproduces (2.5) when the boundary is extended to infinity since then $(R^2/A) \to (1/\pi)$ and $\dot{R} \to 0$. Alike the previous area integrals, Eq. (2.36) can be approximated by a finite sum over galaxy images, but now it extends over a finite area only, thus avoiding boundary effects. Note, however, that $R$ is the distance between the target point $\boldsymbol{x}$ and the boundary via the point $\boldsymbol{x}'$, and $\dot{R}$ is the derivative of this distance with respect to the polar angle around the target point. If $\boldsymbol{x}$ is close to the boundary, $\dot{R}$ can become very large. In addition, if a galaxy happens to be very close to the target point, a small change of the galaxy position can substantially change $R$ and $\dot{R}$. Both these effects can introduce loud noise due to the random positions of the galaxies, if Eq. (2.36) is approximated by a discrete sum over individual galaxy images. It is therefore essential to smooth the data, e.g., according to Eq. (2.12), before inserting them into Eq. (2.36). Even after smoothing, it is not possible to reconstruct $\kappa$ with Eq. (2.36) out to the field boundary; however, simulations show that Eq. (2.36) provides a successful reconstruction mechanism even *close* to the boundary.

Remember that the purpose of this study is to assess the reliability of the total reconstructed cluster mass. Although the finite-field kernel removes the systematic boundary effect from the reconstruction, it yields $\kappa - \bar{\kappa}$ rather than $\kappa$. If we integrated $\kappa - \bar{\kappa}$ out to the field boundary to determine the total enclosed mass, the result would vanish again, by definition of $\bar{\kappa}$, and the only advantage of the finite-field reconstruction would be that the total mass tends to zero not because of systematic errors, but because the underlying $\bar{\kappa}$ of the cluster is unknown. For a mass estimate based on the finite-field reconstruction to be reliable, it is therefore necessary that the mass determination be limited to an area smaller, and preferably much smaller, than the observed field, in order to make $\bar{\kappa}$ much smaller than $\kappa$. In any case, a reconstruction according to Eq. (2.36) provides a lower bound to the total cluster mass, and this bound will become tighter when the observed field is enlarged – see Sect. 3.3 below.

The modified convolution kernel in Eq. (2.36), which I will call "finite-field kernel" in the following, can also be inserted into Eq. (2.23) to replace the kernel previously used in the non-linear reconstruction algorithm. Then, Eq. (2.23) reconstructs the $n$-th iteration of $\kappa - \bar{\kappa}$ instead of $\kappa$ itself, however avoiding systematic boundary effects. If inserted into Eqs. (2.24) and (2.25) to iterate the critical curve and the sign of the Jacobian, this introduces an inaccuracy. In addition, the kernel used in Eq. (2.24) to iterate the shear is still the kernel which applies to the entire real plane instead of the finite field. However, both these inaccuracies are negligible. If we are close to a critical curve, $\kappa \lesssim 1$, hence $\bar{\kappa}$ is much smaller than $\kappa$ in this case if the field is not too small. Furthermore, the only purpose of Eq. (2.34) is to iterate the shear for the determination of $\text{sign}(\det \mathcal{A})$. Since this sign will only change from $+1$ to $-1$ very close to the cluster center, boundary effects introduced by the truncation of the convolution do not matter for that purpose. I will apply this finite-field nonlinear reconstruction to model clusters in the following section.

*2.6 Global statistical measures*

Instead of reconstructing the entire two-dimensional surface-density map of the cluster, the more modest goal of this paper is to estimate masses from weak lensing. A simple way to determine the average convergence enclosed by a boundary starts by using Poisson's equation to write $\kappa$ as the divergence of a vector $\boldsymbol{\alpha} = \nabla \psi$,

$$\kappa = \frac{1}{2} \nabla \cdot \boldsymbol{\alpha} \; ; \tag{2.37}$$

$\boldsymbol{\alpha}$ is the deflection angle caused by the lens. With the two-dimensional version of Gauss' theorem, the area integral over $\kappa$ within a boundary curve $\boldsymbol{c}(\phi)$ can be written

$$\int_A d^2 x \, \kappa(\boldsymbol{x}) = \frac{1}{2} \oint d\phi \, \boldsymbol{\alpha} \cdot \boldsymbol{n} \; , \tag{2.38}$$

where $\boldsymbol{n}$ is the outward-directed vector orthogonal to the boundary's tangent direction, $\boldsymbol{n} = (\dot{c}_2, -\dot{c}_1)$, and thus

$$\int_A d^2 x \, \kappa(\boldsymbol{x}) = \frac{1}{2} \oint d\phi \, (\alpha_1 \dot{c}_2 - \alpha_2 \dot{c}_1) \; . \tag{2.39}$$

Integrating by parts, Eq. (2.39) can be transformed to

$$\int_A d^2 x \, \kappa(\boldsymbol{x}) = \frac{1}{2} \oint d\phi \, (\dot{\alpha}_2 c_1 - \dot{\alpha}_1 c_2) \; . \tag{2.40}$$

If we now specialize $\boldsymbol{c}(\phi)$ to a circle, i.e., $c_1 = R \cos \phi$, $c_2 = R \sin \phi$, we can write

$$\dot{\alpha}_1 = c_1 \alpha_{1,2} - c_2 \alpha_{1,1} \; , \quad \dot{\alpha}_2 = c_1 \alpha_{2,2} - c_2 \alpha_{2,1} \; . \tag{2.41}$$

Moreover, the deflection angle $\boldsymbol{\alpha}$ is the gradient of $\psi$, hence

$$\alpha_{j,k} = \psi_{,jk} = \psi_{,kj} \; , \tag{2.42}$$

and, using Eqs. (2.2) and (2.3),

$$\alpha_{1,1} = \kappa + \gamma_1 \; , \quad \alpha_{2,2} = \kappa - \gamma_1 \; , \quad \alpha_{1,2} = \gamma_2 \; . \tag{2.43}$$

Inserting this together with (2.41) into (2.40), we arrive at the result

$$\bar{\kappa} = \frac{1}{A} \int_A d^2 x \, \kappa = \langle \kappa \rangle + \langle \gamma_{\mathrm{t}} \rangle \; , \tag{2.44}$$

where $\gamma_{\mathrm{t}}$ is the tangential component of the shear along the boundary,

$$\gamma_{\mathrm{t}} \equiv \gamma_1 \cos 2\phi + \gamma_2 \sin 2\phi \; , \tag{2.45}$$

and the angular brackets in Eq. (2.43) denote the average over the boundary curve.

Since $\langle \kappa \rangle \geq 0$ along any boundary, Eq. (2.44) provides a lower limit to the mass enclosed by the boundary,

$$\bar{\kappa}(r) \geq \langle \gamma_{\mathrm{t}} \rangle \; . \tag{2.46}$$

On the other hand, since

$$\bar{\kappa}(r) = \frac{2}{r^2} \int_0^r r' dr' \, \langle \kappa \rangle(r') \; , \tag{2.47}$$

Eq. (2.44) can be transformed to

$$-\frac{2\langle\gamma_{\rm t}\rangle}{r} = \frac{d\bar{\kappa}(r)}{dr} \; . \tag{2.48}$$

Integrating Eq. (2.48) between radii $r_1$ and $r_2 > r_1$, we finally find

$$\zeta(r_1, r_2) \equiv \bar{\kappa}(r_1) - \bar{\kappa}(r_1, r_2) = -\frac{2\,r_2^2}{r_2^2 - r_1^2} \int_{r_1}^{r_2} \frac{dr}{r} \langle\gamma_{\rm t}\rangle \; , \tag{2.49}$$

and the integral can be written as an area integral,

$$\zeta(r_1, r_2) = -\frac{1}{\pi} \frac{r_2^2}{r_2^2 - r_1^2} \int d^2x \, {\rm Re}\left[\mathcal{D}(\boldsymbol{x})\gamma^*(\boldsymbol{x})\right] \; . \tag{2.50}$$

The function $\zeta(r_1, r_2)$ has been derived by Kaiser et al. (1994a), who called it the $\zeta$-statistic. Since $\bar{\kappa}(r_1, r_2)$, the average convergence in an annulus bounded by radii $r_1$ and $r_2$, is non-negative, $\zeta(r_1, r_2)$ provides a lower bound to the average convergence within $r_1$. In principle, Eq. (2.46) could already be used to provide a lower bound on the average convergence within a circular aperture, but it is observationally easier to approximate the area integral in Eq. (2.50) by a sum over all galaxies within a suitably sized annulus and use Kaiser's $\zeta$-statistic as a lower bound on $\bar{\kappa}(r_1)$.

If applied to a set of nested apertures, Kaiser's $\zeta$-statistic can be used to estimate lower bounds on the masses enclosed by the apertures. Consider a set of radii $r_i$, $1 \le i \le n$. If $\zeta$ has been measured in all rings with $r_i < r_j$, Eq. (2.49) can be written

$$\bar{\kappa}_i - \bar{\kappa}_{ij} = \zeta_{ij} \; , \tag{2.51}$$

where $\bar{\kappa}_i \equiv \bar{\kappa}(r_i)$ and $\bar{\kappa}_{ij} \equiv \bar{\kappa}(r_i, r_j)$. The equations (2.51) are not independent since they describe in essence the cluster masses in rings, which must add up to the same mass inside a given radius irrespective of how the radius is decomposed into rings. The masses $M_{ij}$ in rings $r_i \le r \le r_j$ are

$$M_{ij} = A_{ij}\,\bar{\kappa}_{ij} \; , \quad A_{ij} \equiv (r_j^2 - r_i^2)\pi \; , \tag{2.52}$$

and they have to satisfy the relations

$$M_{ij} = \sum_{k=i+1}^{j} M_{k-1,k} \; , \quad M_i = \sum_{k=1}^{i} M_{k-1,k} \; , \tag{2.53}$$

where $M_{0,1} \equiv M_1$. For convenience of notation, we write for the masses inside adjacent rings, $m_k \equiv M_{k-1,k}$. Then, inserting (2.52) and (2.53) into (2.51), we get

$$\frac{A_{ij}}{A_i} \sum_{k=1}^{i} m_k - \sum_{k=i+1}^{j} m_k = A_{ij}\zeta_{ij} \; ; \tag{2.54}$$

a set of $(n/2)(n-1)$ equations for the $n$ unknowns $m_k$. Of these equations, however, all equations for fixed $j$ are redundant, and thus (2.54) provides essentially a set of $n-1$ equations for $n$ unknowns. The fact that there is one equation less than unknowns reflects the scaling invariance of $\kappa$. A reasonable choice is to set $m_n = 0$, i.e., to assume that there is no mass in the outermost aperture. Then, Eqs. (2.54) yield for $m_1$,

$$m_1 = A_1 \frac{A_{1,n}\zeta_{1,n} - A_{1,n-1}\zeta_{1,n-1}}{A_{1,n} - A_{1,n-1}} . \qquad (2.55)$$

Keeping $i = 1$ fixed, Eqs. (2.54) simplify to

$$\sum_{k=2}^{j} m_k = \frac{A_{1,j}}{A_1} m_1 - A_{1,j}\zeta_{1,j} , \qquad (2.56)$$

and the latter set of equations can then be used to determine the masses $M_j$ enclosed by the radii $r_j$,

$$M_j = \left(1 + \frac{A_{1,j}}{A_1}\right) m_1 - A_{1,j}\zeta_{1,j} , \qquad (2.57)$$

for $j \geq 2$. The masses $M_j$ are still a lower bound to the actual masses in circles around the cluster center, because $m_n = 0$ was chosen, but this lower bound will be tighter than $\zeta$ alone.

Note that the restriction to circular boundaries made before Eq. (2.41) simplifies the equations, but is by no means necessary; any shape of a closed boundary would be applicable. In practice, this means that the boundaries chosen could be adapted to the shape of the cluster under consideration, i.e., the boundaries could be shaped like the X-ray isophotes or the galactic isopleths of the cluster.

## 3 Application to cluster models

### 3.1 Sample of model clusters

The numerical cluster models that will be used in the following are taken from the sample described and used in a sequence of earlier papers (Bartelmann & Weiss 1994, Bartelmann, Steinmetz & Weiss 1994, Bartelmann 1994). Briefly, they were produced by $N$-body simulations starting from CDM initial conditions which were normalized to the COBE quadrupole mesurement of the CBR, for the cosmological parameters $\Omega_0 = 1$, $\Lambda = 0$, and $H_0 = 50$ km/s/Mpc. In total, 13 clusters were simulated, and their three-dimensional particle distribution was stored at about ten time steps per cluster between redshifts 1 and 0. The projections of these three-dimensional models along the three independent spatial directions can serve as independent cluster models for the purposes of this paper.

From this numerical cluster sample, I select 60 cluster models with redshifts in the range $z_{\rm cl} \in [0.1, 0.4]$ with a large spread of their maximum convergence $\kappa_{\max} \in [0.4, 1.2]$ with a median of $\bar\kappa_{\max} = 0.8$. The distribution of $\kappa_{\max}$ is displayed as a histogram in Fig. 1.

The clusters are highly irregular in shape, and many of them show substructure and mergers of smaller mass condensations. The fields containing the clusters are square-shaped and have a uniform angular side length of $5'$.

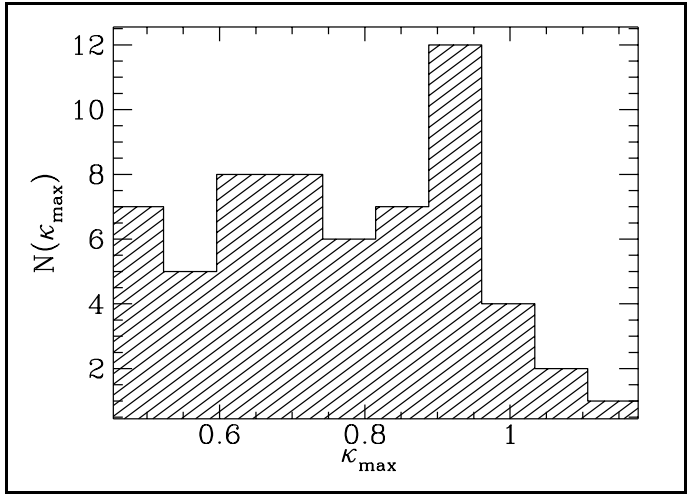

Figure 1.– Histogram of the maximum convergence $\kappa_{\max}$ of the 60 cluster models used in this paper. The median of the distribution is $\bar\kappa_{\max} = 0.8$

### 3.2 Determination of imaging properties

To simulate the distortion of background galaxy images caused by the cluster models, the cluster fields were populated with randomly placed galaxies such that their averaged number density was $n = 35$ (arc min)$^{-1}$, corresponding to an average of $\bar N = 875$ galaxy images per cluster field. Their intrinsic ellipticities $\chi_{\rm s}$ were drawn from a distribution of the form (2.16), with width $\chi_{\rm r} = 0.3$; for this choice, see Miralda-Escudé (1991) and Tyson & Seitzer (1988). With the known convergence and shear fields of the cluster models, the intrinsic ellipticities can be transformed to image ellipticities with

$$\chi = \frac{\chi_{\rm s} + 2g + g^2 \chi_{\rm s}^*}{1 + |g|^2 + 2{\rm Re}(g\chi_{\rm s}^*)} \;, \qquad (3.1)$$

where $g$ is the combination

$$g \equiv \frac{\gamma}{1-\kappa} \;. \qquad (3.2)$$

Once the image ellipticities are found, they can be inserted into the reconstruction equations derived in the previous section.

For each method applied to the cluster sample below, I determine the average reconstructed convergence within circles centered on the cluster center and normalize it by the true average convergence of the input model; the results therefore give the fraction of the reconstructed mass in units of the total mass as a function of distance from the cluster center. The error bars in the following figures show the range containing 80% or 48 of the cluster models, and the curves show the average.

In the following, I apply all the reconstruction methods described above to the numerical cluster sample, viz. the linear reconstruction algorithm without and with smoothing of the input data, the nonlinear iterative reconstruction without and with field extension, the finite-field convolution kernel, either inserted into the linear or the nonlinear reconstruction formulae, without and with field extension, and finally the global mass reconstruction within nested circular apertures, again without and with field extension. Where applicable, the data are smoothed with the adaptive smoothing length given in Eq. (2.18). If the field is synthetically extended, it is extended to twice its side length, i.e., to $10'$. The number of grid cells in the case of the two-dimensional reconstruction techniques is usually $32^2$, which is increased to $64^2$ when the field is extended in order to keep the spatial resolution the same. All the results given below concern the reconstructed mass fraction, i.e., the ratio between the reconstructed mass within a certain radius $M_{\rm rec}(r)$ and the true mass $M_{\rm true}(r)$ within the same radius. For convenience of notation, this ratio is abbreviated by

$$f(r) \equiv \frac{M_{\rm rec}(r)}{M_{\rm true}(r)} \ . \tag{3.3}$$

We can roughly estimate the expected behaviour of $f(r)$. For all reconstruction methods involving infinite integrations, $f(r)$ will go to zero towards the field boundary. For the finite-field reconstructions, assume for simplicity that the cluster is approximately isothermal at sufficient distance from a possible core, hence $\kappa_{\rm true} \propto (1/r)$ and $M_{\rm true} \propto r$. Since $\kappa - \bar\kappa$ is taken as an estimate for $\kappa$ in this case, $f(r)$ will behave like

$$f(r) \simeq 1 - \frac{\bar\kappa r^2 \pi}{M_{\rm true}} \simeq 1 - Cr \ , \tag{3.4}$$

where $C \propto \bar\kappa$; i.e., $f(r)$ will decline linearly from $\simeq 1$ at small $r$ outward.

*3.3 Mass reconstruction from two-dimensional density maps*

As a first example, Fig. 2 displays the result of applying the unsmoothed linear reconstruction formula (2.10) to the simulated image data produced by the numerical cluster sample.

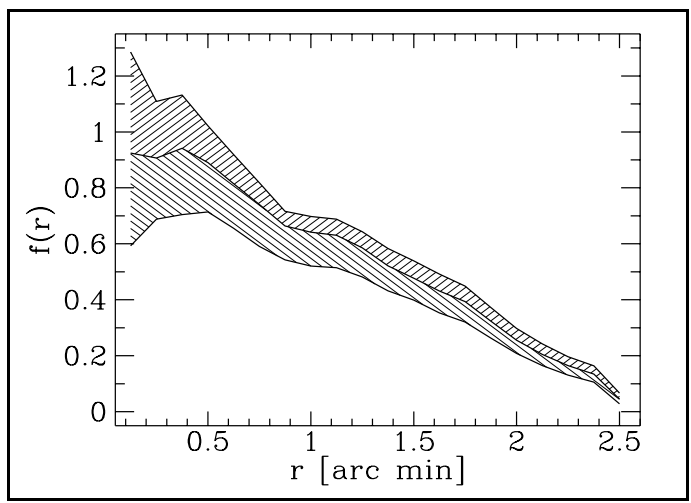

Figure 2.– Linear reconstruction according to Eq. (2.10), without smoothing the input ellipticity data. Close to the cluster center, the mass is reasonably reproduced, but the boundary effects of this method require the total mass within the observed field to vanish, hence the reconstructed mass fraction tends to zero towards the field boundary

The two hatched regions in the figure shows the 80% error interval for the mass reconstruction, together with the average over the sample. It is apparent from the figure that even the most straightforward application of the linear cluster reconstruction algorithm produces reasonable results close to the cluster center; at $r = 0.25'$, $f(r) = 0.90 \pm 0.2$. With increasing radius, however, $f(r)$ tends to zero. The reason for this behaviour is that the applied reconstruction technique approximates a convolution which should extend over the entire real plane by an area integral over a finite field, and therefore the method reconstructs a mass distribution with vanishing total mass within the field. All the more elaborate versions of the reconstruction formula show this behaviour as long as they involve an integration over the entire plane. Panels (a), (b) and (c) of Fig. 3 show examples.

Panel (a) is a reproduction of Fig. 2, added for comparison. Panel (b) shows the result obtained with Eq. (2.14) after smoothing the ellipticity field according to Eq. (2.12). The smoothing of the input data reduces the signal and therefore leads to a lower average $f(r)$ for $r \to 0$, but the scatter in $f(r)$ is reduced; again taken at $r = 0.25'$, $f(r) = 0.73 \pm 0.13$. The nonlinear reconstruction following Eqs. (2.19) through (2.25), the results of which are displayed in panel (c), exhibits the same qualitative behaviour, slightly further reducing the scatter; $f(r) = 0.75 \pm 0.09$ at $r = 0.25'$.

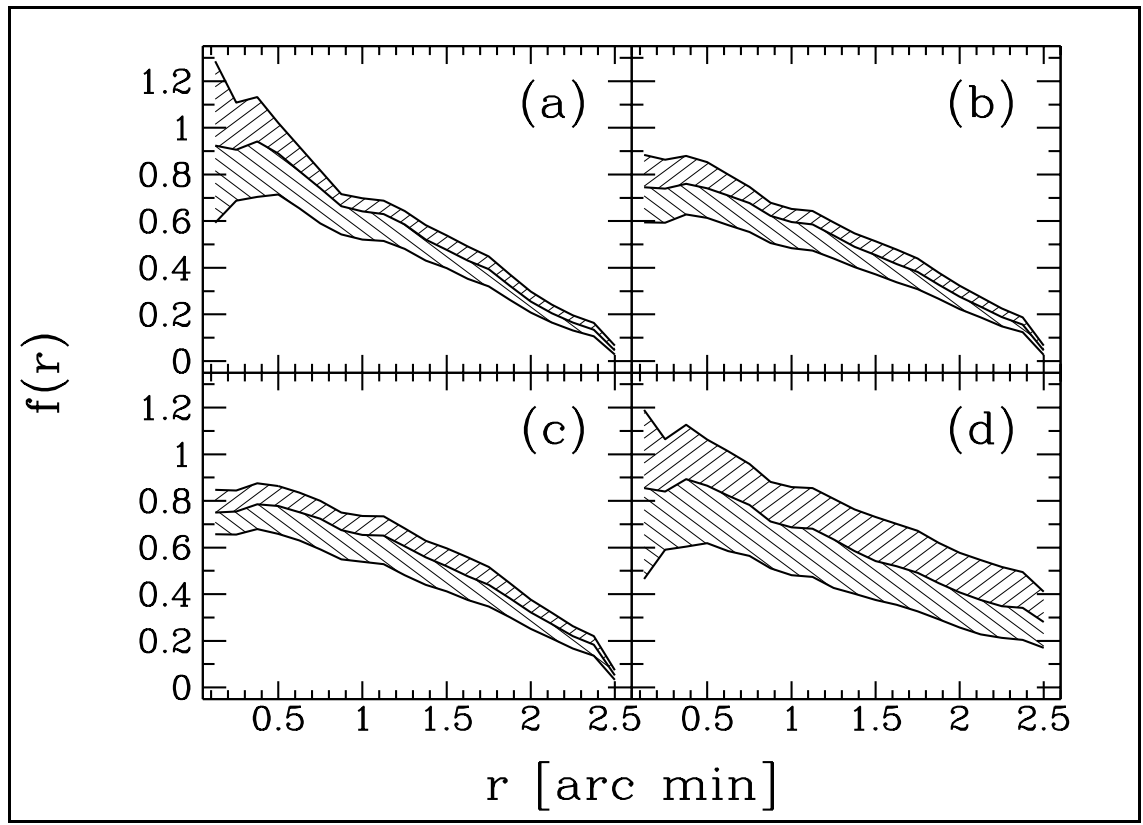

Figure 3.– Comparison of several variants of the reconstruction algorithm. Panel (a): reproduction of Fig. 2, i.e., linear reconstruction without smoothing the ellipticity data; panel (b): linear reconstruction after smoothing the ellipticity data; panel (c): nonlinear reconstruction; panel (d): reconstruction with the finite-field kernel of Eq. (2.36). From (a) to (b), part of the signal is lost due to smoothing, but the errors towards the center decrease. The nonlinear reconstruction (c) further reduces the error close to the center. Panels (a) through (c) clearly show that the boundary effects require a vanishing total mass in the field. The finite-field reconstruction (d) removes this boundary effect, but increases the error

Besides the unwanted feature of reconstructing a vanishing total mass, these results show that the error in the straightforward linear reconstruction formula (2.10) can be re-duced by about a third when the input data field is smoothed prior to the convolution, and by about a half when the additional modifications required by the nonlinear reconstruction algorithm are included. The amount of the latter reduction, which comes from a more reliable reconstruction of the clusters with high $\kappa_{\max}$ in our numerical sample, depends of course on the distribution of peak values for $\kappa$ in the actual cluster sample. For comparison, panel (d) of Fig. 3 shows the result of the finite-field reconstruction following Eq. (2.36), i.e., a reconstruction which does not involve an infinite integration. As was to be expected, the mass-reconstruction function $f(r)$ does no longer tend to zero when $r$ approaches the field boundary, but remains finite.

One could attempt to employ the scaling invariance of the distortion field $\delta$ to rescale the convergence field by the global transformation

$$\kappa \to \kappa' = (1-\lambda) + \lambda\kappa , \qquad (3.5)$$

in order to correct for the systematic boundary effect which forces the total mass in the field to vanish. Since the decay of $f(r)$ for $r$ going towards the field boundary arises in part because the simpler variants of the reconstruction method create spurious negative troughs in the convergence field, a plausible rescaling of $\kappa$ would be to choose $\lambda$ such that $\kappa$ is non-negative everywhere. As an example, we display in Fig. 4 the results of such a rescaling of the convergence maps obtained from the linear reconstruction equation (2.10).

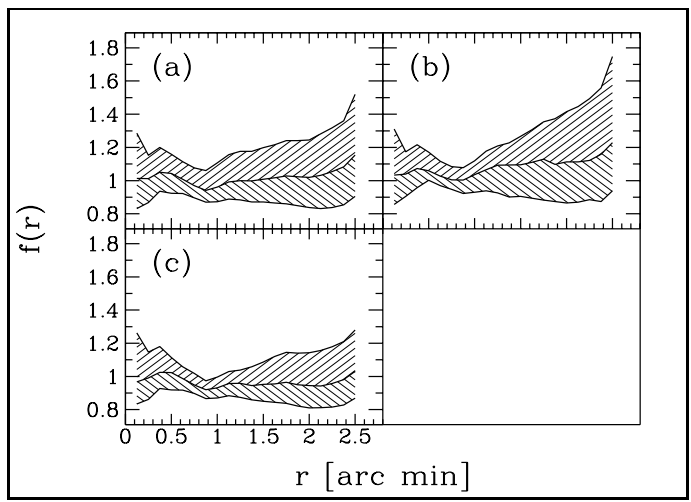

Figure 4.– Linear reconstruction without smoothing the ellipticity data, scaled with the global transformation (3.2) such that $\kappa \geq 0$ everywhere in the field. Panel (a): entire sample; panel (b): $\kappa_{\max} \leq 0.8$, panel (c): $\kappa_{\max} > 0.8$. While the rescaling yields reasonable results for the high-$\kappa_{\max}$ subsample, it tends to substantially overestimate the total mass for large radii

Panel (a) of the figure shows the results for the entire cluster sample, panel (b) for the low-$\kappa$ clusters ($\kappa_{\max} \leq 0.8$), and panel (c) for the high-$\kappa$ clusters. While the results for the high-$\kappa$ clusters are fairly accurate, $f(r) \in [0.87\ldots 1.28]$ at $r = 2.5'$, the errors for the low-$\kappa$ clusters become substantial at this radius, $f(r) = [0.94\ldots 1.75]$. The reason is that this rescaling procedure exaggerates the necessary correction. If the signal, i.e., the shear-induced ellipticities of the images becomes small compared to the intrinsic width of the source ellipticity distribution, substantial fluctuations in the convergence map arise which are not systematic like the spurious negative troughs, but random. Since this noise has a white spectrum, positive and negative fluctuations of that kind should cancel on average, but the rescaling described here lifts the convergence map so far that $\kappa \geq 0$ everywhere.

Therefore, positive fluctuations in $\kappa$ can no longer cancel out, and the rescaled convergence map overestimates the total mass. This effect is strongest for low-$\kappa$ clusters since there the overall signal is weaker compared to the intrinsic ellipticity.

As mentioned before, the convolution kernel of Eq. (2.36), which was derived to avoid systematic boundary effects, can straightforwardly be inserted into the nonlinear reconstruction algorithm, which should result in a better mass reconstruction for small radii. Panel (a) of Fig. 5 shows the results obtained with that technique.

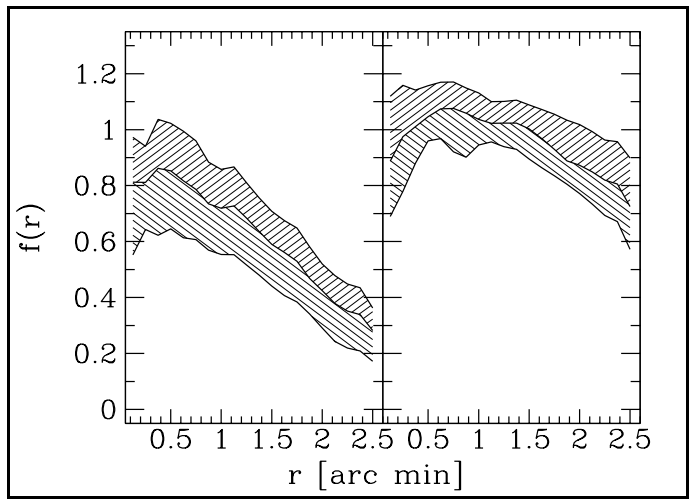

Figure 5.– Nonlinear reconstruction using the finite-field convolution kernel of Eq. (2.36). Panel (a): using only data in the original field; panel (b): after extending the field synthetically. The nonlinear reconstruction reduces the error close to the cluster center (cf. panel (d) of Fig. 3), and the field extension improves the reconstruction result considerably

As can be seen by comparing panel (a) of Fig. 5 to panel (d) of Fig. 3, the nonlinear reconstruction algorithm reduces the error for small $r$, while leaving the results for larger $r$ basically unchanged. Taken again at $r = 0.25'$, $f(r) \in [0.59\ldots 1.1]$ for the linear and $f(r) \in [0.64\ldots 0.94]$ for the nonlinear reconstruction.

Although Eq. (2.36) avoids the unwanted negative troughs, it has the disadvantage that it does not reconstruct the convergence field, but rather $\kappa(\boldsymbol{x}) - \bar{\kappa}$, i.e., the convergence reduced by $\kappa$ averaged within the field. (Note that this is *not* a transformation of the type (3.2).) It therefore seems promising to reduce $\bar{\kappa}$ by extending the field. In panel (b) of Fig. 5, I show the results for $f(r)$ obtained in the following way. First, the data field is extended according to the prescription (2.26), i.e., by adapting a spherically symmetric, parametrized lens model to the distortion data and using this lens model then for creating synthetic distortion data outside the 'observed' field. Then, the nonlinear reconstruction technique with the finite-field convolution kernel of Eq. (2.36) is applied to the extended $\delta$ field. Especially in comparison with panel (a) of Fig. 5, panel (b) shows that this procedure is successful; it not only reduces the noise of the mass reconstruction but also corrects for part of the decline for increasing $r$ apparent in panel (a).

### 3.4 Lower bounds to the radial mass profile

For the sole purpose of reconstructing masses within apertures, the construction of two-dimensional mass maps is not necessary. I therefore also apply Eq. (2.57) to the synthetic image data; the results are displayed in Fig. 6.

The comparably small error bars apparent in Fig. 6 are due to the fact that, in determining the $\zeta_{1,j}$ required in Eq. (2.57), the image data are averaged within much larger

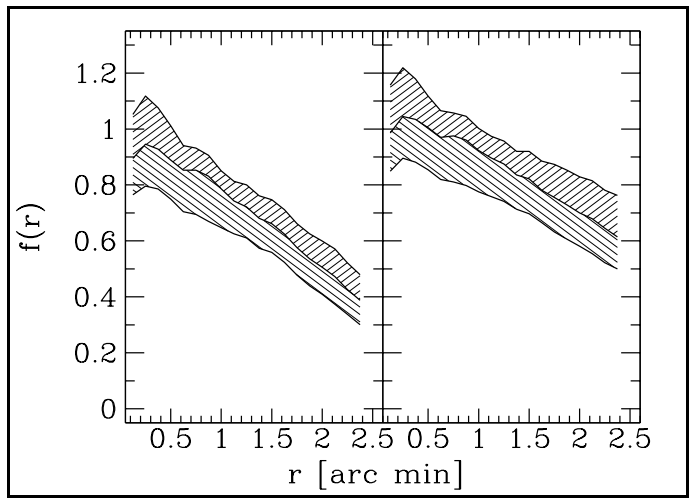

Figure 6.– Lower bounds to the mass enclosed in circular apertures. Panel (a): using only the observed field; panel (b): after synthetically extending the field. Especially after field extension, the mass reconstruction close to the cluster center is very good, but it drops to $\simeq 60\%$ of the mass close to the field boundary

parts of the observed field. Since the boundary condition $m_n = 0$ entered the derivation of Eq. (2.57), the result is a lower bound to the actual mass. It underestimates the true mass by more than 50% close to the boundary of the field; see panel (a). Again, the results improve if the data field is extended with the help of a parametrized, symmetric lens model. Panel (b) of Fig. (6) shows the result for the lower bound to the radial mass profile according to Eq. (2.57), obtained after extending the distortion field. The field extension increases the reconstructed mass fraction from 40% to 60% close to the field boundary, and it also improves the accuracy of the reconstruction towards the center of the field. At $r = 0.25'$, $f(r) = 0.95 \pm 0.15$ for the original data field, and $f(r) = 1.0 \pm 0.1$ for the extended data field.

## 4 Influence of seeing and the source redshift distribution

Seeing reduces the signal because it tends to circularize images. The effect of seeing can be approximately quantified in the following way. Suppose the surface-brightness distribution of the source is Gaussian,

$$I(\boldsymbol{x}) \propto \exp\left(-\frac{x_1^2}{a^2} - \frac{x_2^2}{b^2}\right) , \qquad (4.1)$$

where $a$ and $b$ are the semi-major and semi-minor axes of the image ellipse. The effect of seeing is modeled by convolving $I(\boldsymbol{x})$ with a Gaussian of width $s$. It is then straightforward to show, most easily with the convolution theorem for Fourier transforms, that the ellipticity $|\chi|'$ of the image affected by seeing is related to $|\chi|$ by

$$|\chi|' = \frac{|\chi|}{1 + \sigma^2(1 - |\chi|)} , \qquad (4.2)$$

where $\sigma$ is the width of the seeing disk in units of the intrinsic source width $b$, $\sigma \equiv (s/b)$. Since the reconstructed convergence is linear in $\chi$ within the largest fraction of the observed

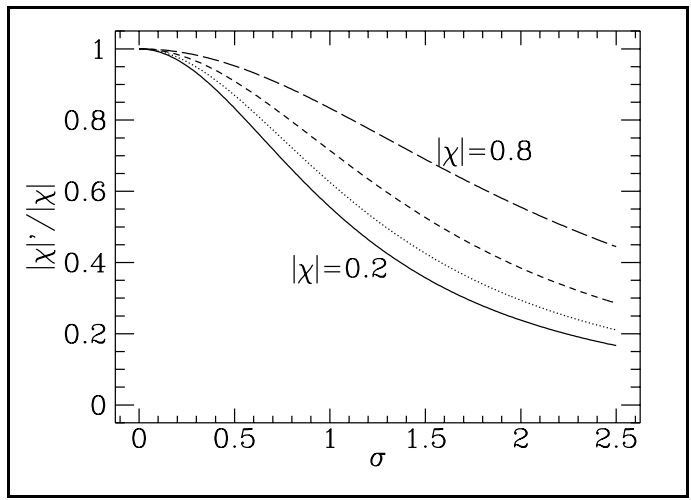

Figure 7.– Influence of seeing on the ellipticity of galaxy images; their surface-brightness profile was assumed Gaussian. The figure shows the ellipticity after seeing, divided by the true ellipticity, as a function of the width of the seeing disk in units of source widths, $\sigma = (s/b)$. The effect of seeing becomes stronger for weaker signal, as expected. Signals of moderate amplitude, $|\chi| \simeq 0.5$ say, are reduced by $\simeq 50\%$ if $\sigma \simeq 1.4$

cluster field, Eq. (4.2) gives an estimate for the effect of seeing on the surface-density maps. Fig. 7 displays $(|\chi|'/|\chi|)$ in dependence of $\sigma$ for four different values of $|\chi|$.

Clearly, a very strong signal, $|\chi| \to 1$, is hardly affected by seeing at all; it requires a width of the seeing disk of

$$\sigma \simeq \sqrt{\frac{1}{1-|\chi|}} \tag{4.3}$$

in order to reduce the signal by 50%. A signal of moderate strength, say $|\chi| = 0.5\ldots 0.6$, is reduced to one half if the seeing is roughly 1.5 times the source size. In principle, it is possible to correct for the decrease of the signal due to seeing using Eq. (4.2).

Another systematic source of error is the unknown redshift distribution of the sources. Assuming a certain source redshift $z'_s$ different from their true redshift $z_s$ leads to an overestimate of the cluster mass if $z_s > z'_s$, and conversely. This can be quantified by the geometrical factor entering the conversion between surface mass density and convergence,

$$\kappa(z_d, z_s) \propto \frac{D_d D_{ds}}{D_s}, \tag{4.4}$$

where $D_{d,s,ds}$ are angular-diameter distances from the observer to the cluster and to the source, and from the cluster to the source, respectively. Fig. 8 displays the function

$$r(z_d; z_s, z'_s) \equiv \frac{\kappa(z_d, z_s)}{\kappa(z_d, z'_s)} = \frac{D_{ds}}{D_s} \frac{D'_s}{D'_{ds}}, \tag{4.5}$$

i.e., the ratio between the convergences for the true and the hypothesized source redshift, respectively. The definition implies that $r$ is the factor by which the cluster mass is *overestimated* if $z'_s$ is assumed in place of the true source redshift. For the figure, $z'_s = 1$ is kept fixed, and curves for four cluster redshifts $z_d \in \{0.1, 0.2, 0.3, 0.4\}$ are shown.

The figure shows that the uncertainty due to the unknown source redshifts is negligible if the cluster is at a redshift significantly lower than the sources; for $z_d = 0.2$, $r \in [0.89, 1.06]$ for $z_s \in [0.6, 2.0]$. For high-redshift clusters the error can be substantial; $r \in [0.33, 1.34]$ for $z_d = 0.5$. For moderate-redshift clusters, however, the uncertainty in the source redshifts translates to an uncertainty in the cluster mass on the order of $10\%\ldots 20\%$. If the sources are distributed in redshift, $r$ has to be integrated over the source-redshift interval, weighted by the redshift distribution function. The curves in Fig. 8 were calculated for the parameters of an Einstein-de Sitter universe, but they are very insensitive to changes in the cosmological model.

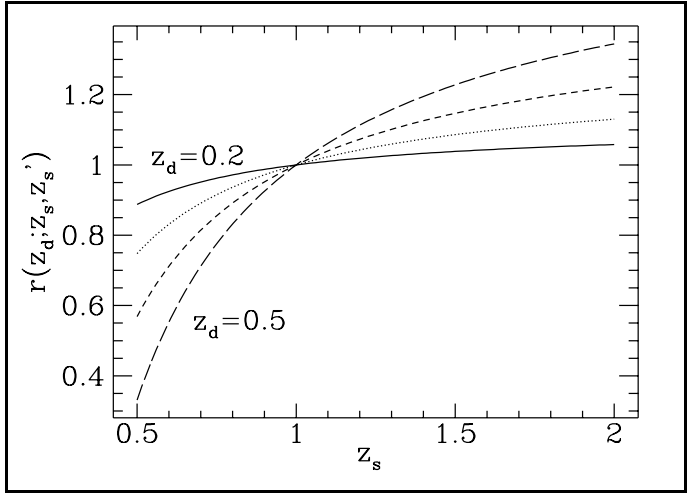

Figure 8.– Uncertainty caused by the unknown source redshift distribution, quantified by the function $r$ defined in Eq. (4.5); $r$ is the factor by which the cluster mass is *overestimated* if the sources are assumed at redshift $z'_s = 1$ while their true redshift is $z_s$. The result depends on the cluster redshift $z_d$ as indicated; for clusters at low redshifts, the uncertainty is negligible, and it increases for increasing $z_d$. $r$ is very insensitive to the cosmological parameters $\Omega_0$ and $\Lambda_0$; for the plot, an Einstein-de Sitter model was assumed

## 5 Summary

A number of different cluster reconstruction techniques based on weak lensing were reviewed and applied to a sample of numerically simulated cluster mass distributions. The following methods were discussed:

1. Linear reconstruction without smoothing the observed galaxy ellipticities (Kaiser & Squires 1993). This method is designed for weak lens effects, i.e., for $\kappa \ll 1$ and $\gamma \ll 1$. It involves low-pass filtering of the signal to reduce the otherwise infinite noise due to the random ellipticities and positions of the galaxies.
2. Linear reconstruction after smoothing the galaxy ellipticities (Seitz & Schneider 1994). This method is a variant of (1.) and still operates in the weak-lensing regime.
3. Non-linear iterative reconstruction (Seitz & Schneider 1994). This technique proceeds beyond the weak-lensing limit, and therefore has to control the critical curves of the cluster.
4. Finite-field reconstruction (e.g., Schneider 1994, Kaiser et al. 1994c), which replaces the convolution kernel operating on $\mathbb{R}^2$ required by methods (1.) through (3.) to convert shear into convergence by a kernel which operates on a finite field only. This replacement avoids boundary effects which reduce the total mass within the observed field to zero. This finite-field kernel can be used by methods (2.) and (3.), regardless of non-linearities in the lens effect.
5. Synthetic field extension (Seitz & Schneider 1994), which was suggested as another means to reduce the boundary effects of methods (1.) through (3.). It proceeds by adapting a spherically symmetric lens model, parametrized by central convergence, core radius, and convergence-profile slope, to the outer parts of the observed ellipticity data, and then synthetically extending the measurements beyond the field boundaries.

   Again, this field extension is independent of the reconstruction technique chosen and
   can be combined with either.

6. Mass reconstruction within circular apertures (based on the $\zeta$-statistic by Kaiser et
   al. 1994a). This method employs the fact that, by Gauss' theorem in the plane,
   the average convergence within a boundary can be expressed by a line integral over
   observable quantities along the boundary.

The results described above can be summarized in the following way.

a. The accuracy of the mass reconstruction close to the cluster center is acceptable even
   for the most straightforward application of the linear reconstruction algorithm. The
   error in the function $f(r)$ defined in Eq. (3.3) is $\pm 0.2\%$ for method (1.), and can be
   reduced by methods (2.) and (3.) to $\pm 0.13$ and $\pm 0.09$, respectively. The improvement
   achieved by method (2.) is a result of smoothing the observed data prior to use, and
   the further improvement by method (3.) is due to the more accurate reconstruction
   of the centers of clusters with central convergence $\kappa_{\max} \simeq 1$.

b. A substantial disadvantage of methods (1.) through (3.) is that they approximate an
   integral which should cover $\mathbb{R}^2$ by an integral over the observed field only. As a result,
   they reconstruct a mass distribution in the field with vanishing total mass by producing
   spurious negative troughs in the convergence map close to the field boundary. A
   possible cure would be to employ the global scaling invariance (3.5) to scale $\kappa$ such
   that it is non-negative everywhere in the field. This, however, removes some negative
   fluctuations in $\kappa$ arising from the statistical fluctuations in the intrinsic ellipticities
   of the sources which should in the ideal case cancel against positive fluctuations.
   Correcting such that $\kappa \geq 0$ everywhere in the field thus overestimates the cluster
   mass especially if the cluster's lens effect is weak, because if $\kappa$ is small, the statistical
   fluctuations are relatively stronger compared to the true $\kappa$ than if the lens effect is
   stronger. While the rescaling to non-negative $\kappa$ works quite well for clusters with high
   central convergence, $\kappa_{\max} \gtrsim 0.8$ say, it can produce a substantial overestimate of the
   cluster mass, by up to a factor of $\simeq 1.8$ in our simulations, depending on the size of
   the field.

c. The method to correct for this boundary effect employed here, based on Eq. (2.36), is
   successful, but it suffers from a number of different shortcomings. First, it does not
   reconstruct $\kappa$ directly, but rather the difference $\kappa - \bar{\kappa}$, i.e., the convergence reduced
   by the average convergence in the field. However, $\bar{\kappa}$ can be reduced by enlarging
   the field. Second, the finite-field convolution kernel of Eq. (2.36) is susceptible to
   noise especially close to the field boundary and if unsmoothed observational data
   are used. It is therefore necessary to smooth the data, e.g., according to Eq. (2.12),
   before inserting them into Eq. (2.36), and to avoid a region close to the field boundary
   which can, however, be chosen small. If these precautions are taken, the finite-field
   reconstruction is promising, especially since it can be combined with the non-linear
   reconstruction method.

d. Further substantial improvement of the quality of the mass reconstruction is achieved
   by the synthetic field extension. The reason for this is simply that it reduces $\bar{\kappa}$ in
   the effective field, and thus raises the lower bound on $\kappa$ provided by the finite-field
   reconstruction. As can be read off from Fig. 5, the uncertainty of the finite-field mass reconstruction relative to the average after field extension is $\simeq 0.1$, and even close to the boundary the average reconstructed mass fraction is $\simeq 0.8$.

e. The mass reconstruction within a set of nested apertures, which also underestimates the true mass, performs equally well as the non-linear finite-field reconstruction after field extension in terms of its uncertainty, but it provides a less tight lower bound than the latter.

On the whole, two main conclusions can be drawn. First, cluster reconstruction techniques which require a convolution integral to be performed over the entire real plane do not yield reliable mass reconstructions, because they introduce boundary effects which are hard to control. Overestimates of the mass are more likely and more substantial than underestimates, especially if the lens effect of the cluster under consideration is weak. Second, modified convolution kernels which apply to a finite field of integration yield lower bounds to the total mass. The tightness of these bounds depend on the size of the field. For the side length of $5'$ of the fields numerically simulated here, the reconstructed mass fraction $f(r)$ drops from $\simeq 0.8$ at the cluster center to $\simeq 0.3$ at a distance of $2\rlap.'5$, and this can be improved by synthetic field extension to $f(r) \simeq 1.0$ at $r = 0$ and $f(r) \simeq 0.8$ at $r = 2\rlap.'5$. Based on these results, it appears probable that the cluster reconstruction techniques based on the weak lens effect can achieve an accuracy of the reconstructed mass on the order of $10\% \ldots 20\%$, if they are applied to fields with side lengths of $10'$ or larger.

Some cautionary remarks must finally be added. First, seeing deteriorates the intrinsic signal, and guiding errors, telesope aberrations and similar effects can cause spurious signals. However, the exceptional quality of data taken with the refurbished HST (as an example, see plate L11 of Dressler et al. 1994), combined with very elaborate techniques of data analysis (cf. Kaiser et al. 1994b), give rise to the expectation that these difficulties can be overcome. Second, the redshift distribution of the arclet sources is yet unknown, and this introduces an uncertainty which can become substantial for high-redshift clusters. Third, all methods described here suffer from the scaling invariance (3.5) of the convergence, which leaves the observable signal unchanged. An additional sheet of constant surface density would therefore never be detectable with methods based on shear alone. This degeneracy can be removed by either requiring that $\kappa \to 0$ outward, or observationally by methods which are sensitive to the magnification of images. Fourth, clusters are embedded into large-scale structures, which by themselves contribute shear and magnification on a level of $10\% \ldots 20\%$. Possible effects of these additional large-scale lenses will be the subject of a further study.